\documentstyle[floats,psfig,aps]{revtex}

\newcommand\jps{{J/\psi}}
\newcommand\ccb{{c\bar{c}}}

\begin{document}

\twocolumn
\title{Charmonium suppression in p-A collisions at RHIC}

\author{Fran\c{c}ois Arleo, Pol-Bernard Gossiaux, and J\"org Aichelin}

\address{SUBATECH \\
Laboratoire de Physique Subatomique et des Technologies Associ\'ees\\
UMR Universit\'e de Nantes, IN2P3/CNRS, Ecole des Mines de Nantes\\
4, rue Alfred Kastler,
F-44070 Nantes Cedex 03, France.}
\author{\begin{quote}
\begin{abstract}
We discuss charmonium production in proton-nucleus collisions at RHIC energies
under the assumption of $x_F$ and $x_2$ scaling. We find that all the
ambiguities due to energy loss are gone at this energy and therefore data will
reveal the scaling law, if any. These p-A data will also be crucial to interpret
nucleus-nucleus data with respect to a possible formation of a quark gluon
plasma because the extrapolations for charmonium production from the present p-A
data to RHIC energies, based on the two scaling laws, differ by a factor of
four. 
\end{abstract}
\pacs{~}
\end{quote}}
\maketitle

\section{Scaling in charmonium nuclear production} 

Charmonium production off nuclei has shown to be strongly affected by nuclear effects in proton induced reactions at several beam energies~\cite{na3,na38,e772,e866}. It is therefore natural to wonder whether the nuclear dependence of charmonium production is actually driven by a scaling variable, independently of the center-of-mass energy $\sqrt{s}$ of the collision. Indeed, a first glimpse at proton-nucleus data at SPS~\cite{na3} and Fermilab~\cite{e772,e866} tends to indicate that $J/\psi$ suppression scales with the longitudinal momentum fraction Feynman-$x$ ($x_F$).

Such an $x_F$ scaling has been predicted long ago within the intrinsic charm model~\cite{icm}. Indeed, charmonium production at large $x_F$ may be dominated by the soft scattering of the intrinsic $c\bar{c}$ component of the projectile on the front surface of the nucleus leading to an unusual $A^{2/3}$ dependence at high $x_F$, i.e., a suppression as compared to the $A^1$ dependence of the hard parton fusion process. However, the intrinsic charm content of the proton required to describe quantitatively the data is somewhat larger than what is found from a leading order analysis of the EMC charm structure functions~\cite{Hoffmann}.

Furthermore, it is expected from the QCD factorization theorem that nuclear effects at high energy scale with the momentum fraction $x_2$ carried by the target parton from the nucleus. Furthermore, absorption of charmonia in nuclear matter depends on the space-time evolution of the expanding $\ccb$ state~\cite{ccbevol} as well as on the energy dependence of the charmonium hadron inelastic cross sections~\cite{energydep}. Since both the Lorentz factor $\gamma$ of the $\ccb$ system in the nucleus rest frame (i.e., which governs the expansion of the $\ccb$ precursor in the nucleus) and the incident energy $\sqrt{s_{\psi h}}$ are related to the momentum fraction $x_2$, final state interactions naturally lead to an $x_2$ scaling in charmonium suppression, assuming a given mass for the $\ccb$ state. Similarly, the effects of nuclear shadowing~\cite{shadowing} in charmonium production are expected to depend on $x_2$, independently of the center-of-mass energy of the collision. However, we would like to point out that possible effects of parton energy loss in nuclear matter might distort a possible $x_2$ scaling law at low beam energy~\cite{moriond00,arleo_these}.

On the experimental side, we emphasized that the comparison of NA3~\cite{na3} ($\sqrt{s} = 19.4$~GeV) and E772~\cite{e772} ($\sqrt{s} = 38.8$~GeV) data seems to show that $J/\psi$ suppression in proton-nucleus collisions scales with the longitudinal momentum fraction Feynman-$x$ ($x_F$)~\cite{e772}. However, $x_F$ scaling is no longer reported in pion-nucleus NA3 data~\cite{na3} from $\sqrt{s} = 16.7$~GeV to $\sqrt{s} = 22.3$~GeV~\cite{gup92}. The $x_2$ scaling assumption appears to be ruled out at small $x_2 \approx 0.07$ when considering NA3 and E866/NuSea p-A data, even though a good agreement is found at larger $x_2$. Thus, it is our understanding that two quite different scenarios can be invoked in order to explain the present p-A data : i) A universal $x_F$ scaling law, valid from $\sqrt{s}\approx 20$ GeV on; ii) A $x_2$ scaling law visible at higher energy, but distorded in the low energy range by effects such as energy loss. Unfortunately, no clear and definitive conclusion can be drawn from the present experimental results, first of all because considerable error bars weaken the confidence one could have in some of the most relevant data points, but also because energy loss effect is in fact not perfectly understood at NA3 energies. 

In order to distinguish clearly which scaling law, if any, governs nuclear effects seen in charmonium production, one thus has to turn to higher $\sqrt{s}$, such as presently available at the RHIC accelerator. It is the main goal of this article to show that energy loss effects are indeed negligible in this energy range, while $x_2$ and $x_F$ scaling law assumptions lead to quite different predictions for the production rate of charmonium in proton-nucleus collisions planned in year III of RHIC. Consequences for the interpretation of nucleus-nucleus experiment are eventually investigated. 

\section{RHIC predictions} 

Using the 800~GeV proton beam at Fermilab, the
fixed target experiment E866/NuSea measured $J/\psi$ and $\psi'$ production off
several nuclear targets (Be, Fe, and W) over a wide kinematical range~
\cite{e866}. The production ratio
\[ 
R({\mathrm W}/{\mathrm Be}, x_F)=\frac{9\,{\rm d}\sigma(p \,{\rm W}\, \to \psi(x_F)\, X)}{184\,{\rm d}\sigma(p \,{\rm Be}\, \to \psi(x_F)\, X)}, 
\]
has been extracted for longitudinal momentum fractions $-0.1 \le x_F \le 0.93$.
It is found to decrease strongly from $R$(W/Be)=0.86 at $x_F \approx 0.2$ down
to $R$(W/Be)=0.3 at $x_F=0.9$. 

In~\cite{e866}, the $x_F$ dependence of the differential production ratio has
shown to be consistent with 
\[
R({\mathrm W}/{\mathrm Be},x_F) = \left(\frac{184}{9}\right)^{\alpha(x_F)-1}
\]
where $\alpha(x_F)$ follows the fitting law
\[
\alpha^{E866}(x_F) = 0.960\,\left(1-0.0519\,x_F - 0.338\,x_F^2\right)
\]
in the whole experimental $x_F$ range.

\begin{figure}[htbp] 
\centerline{\psfig{figure=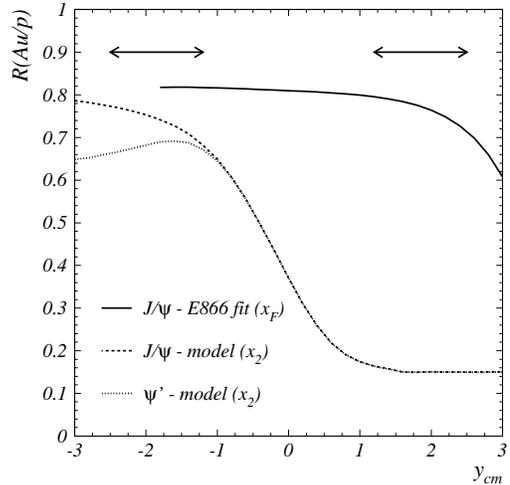,width=7.cm}}
\caption{Ratio R(Au/p) versus $y_{\mathrm{cm}}$ of produced $J/\psi$'s within
the E866/NuSea fit (solid line) in p(100~GeV)+ Au(100~GeV) collisions, assuming
$x_{F}$ scaling. The model prediction for $J/\psi$ (dashed) and $\psi'$ (dotted)
production ratio is also shown. The arrows show the dimuon arm acceptance in the
PHENIX experiment.} 
\label{nuau} 
\end{figure}

\emph{$x_F$ scaling}. Assuming $x_F$ to be the proper scaling variable, i.e.,
$R$ independent of $\sqrt s$, the expected $J/\psi$ production rate at RHIC is
straightforward. Provided the power law $\sigma(A)~\propto~A^\alpha$ is valid
even for light systems, the ratio $R({\mathrm Au}/{\mathrm p},x_F)$ in p(100
GeV)+A(100 AGeV) is simply given by
\[
R({\mathrm Au}/{\mathrm p},x_F) = 197^{\alpha^{E866}(x_F)-1}.
\]
The simple relation between $x_F$ and the $J/\psi$ rapidity
\[
y^{J/\psi} = {\mathrm Arcsinh} \left( \frac{x_F \sqrt{s}}{2 \,m_{J/\psi}} 
\right)\,,
\]
determines the rapidity range over which reliable RHIC predictions, based on
E866 data, can be made. This range is summarized in
Table~\ref{table_xf_scaling}. The corresponding production ratio is displayed in
Figure~\ref{nuau} ($solid$) as a function of $y^{J/\psi}_{\rm{cm}}$.
Provided $x_F$ scaling, charmonium suppression is expected to be small ($R
\approx 0.8$) and approximately constant in the range $-2 \le
y^{J/\psi}_{\rm{cm}} \le 2$. The large nuclear effects seen in E866 data are
found to be relevant at RHIC for rapidities greater than
$y^{J/\psi}_{\rm{cm}} \approx$ 2.6 (cf. Table~\ref{table_xf_scaling}, {\it
right} and Figure~\ref{nuau}).
\begin{table}
\begin{center}
\begin{tabular}{|p{2.3cm}|p{3.cm}|p{3.cm}|}
 & \centerline{full range} & \centerline{steep drop} \\
\hline 
$x_{F}$ in E866 & \centerline{$-0.1 \le x_{F}\le 0.93$} & \centerline{$0.2\le
x_{F}\le 0.93$}  \\
\hline
$y_{\rm cm}^{J/\psi}$ at RHIC & \centerline{$-1.9\le y_{\rm cm}^{J/\psi}\le
4.1$} & 
\centerline{$2.6\le y_{\rm cm}^{J/\psi} \le 4.1$} 
\end{tabular}
\end{center}
\caption{$x_F$ range covered by the E866/NuSea data and the corresponding
$J/\psi$ rapidity range at RHIC (${\sqrt s}$~=~200 AGeV) ($left$). Ranges for
which the ratio R(W/Be) sharply drops are also given ($right$).}
\label{table_xf_scaling}
\end{table}

\emph{$x_2$ Scaling}. The scaling with the momentum fraction $x_2$ leads to a
significantly different suppression pattern. Expressing $x_2$ as a function of
$x_F$
\[ 
x_2^{J/\psi} = \frac{1}{2}\left(- x_F + \sqrt{ x_F^2 + 4 m_{J/\psi}^2/s}
\right) 
\] 
and the $J/\psi$ rapidity as a function of $x_{2}$  
\[ y^{J/\psi}_{\mathrm{cm}} = {\mathrm ln} \left(\frac{m_{J/\psi}}{x_2^{J/\psi} 
\sqrt{s}}\right)\,,
\] 
the sudden drop of the production ratio R occurs around midrapidity at RHIC
energies, as shown in Table~\ref{table_x2_scaling} ({\it right}). Furthermore,
we note that $x_2$ scaling predictions, based on E866 data, are limited to the
rapidity range  $-1.1 \le y^{J/\psi}_{\mathrm{cm}} \le 0.8$
(Table~\ref{table_x2_scaling}, {\it left}).
\begin{table}[htbp]
\begin{center}
\begin{tabular}{|p{2.3cm}|p{3.cm}|p{3.cm}|}
 & \centerline{full range} & \centerline{steep drop} \\
\hline
$x_{2}$ in E866 & \centerline{$0.044\ge x_{2}\ge 0.006$} & \centerline{$0.028\ge
x_{2}\ge 0.006$}\\
\hline
$y_{\rm cm}^{J/\psi}$ at RHIC & \centerline{$-1.1\le y_{\rm cm}^{J/\psi} \le
0.8$} & 
\centerline{$-0.6\le y_{\rm cm}^{J/\psi} \le 0.8$}
\end{tabular}
\end{center}
\caption{$x_2$ range covered by the E866/NuSea data and the corresponding
$J/\psi$ rapidity range at RHIC (${\sqrt s}$~=~200 AGeV) ($left$). Ranges for
which the ratio R(W/Be) sharply drops are also given ($right$).}
\label{table_x2_scaling}
\end{table}

Thus, the steep fall of charmonium production seen
in E866 data appears in two distinct rapidity ranges at RHIC, depending on whether
it scales with $x_F$ or $x_2$. Consequently, the rapidity dependence of
charmonium production measured in proton-nucleus collisions at RHIC may allow
one to disentangle these two assumptions. 

\emph{The Model}. The expected charmonium rate at RHIC under the $x_2$ scaling
assumption is discussed in this section. In order to extrapolate to $x_2$ ranges
not covered by E866 data (Table~\ref{table_x2_scaling}), we use a recently
advanced $x_2$ scaling model based on absorption mechanisms~\cite{arl99}. Here,
it will be complemented by a calculation of parton energy loss, that has been
proven to be relevant at low beam energy~\cite{moriond00}.

The model is based on a standard two stage color neutralization scenario. A
compact partonic $(c\bar{c})_8$ state is assumed to be produced by parton fusion
in a color octet state. The latter will then turn into a singlet $(c\bar{c})_1$
state after a color neutralization time $\tau_{8 \rightarrow 1}$. During the
first stage, the octet state may interact with surrounding nucleons with an
octet cross section $\sigma_{(c\bar{c})_8N}$, taken to be independent of its
transverse size $r_{c\bar{c}}$. Adopting an energy dependence related to
$J/\psi$ photoproduction data, the octet
nucleon cross section reads as
\[ 
\sigma_{(c\bar{c})_8N}=\sigma_{8}\cdot \left(\frac{\sqrt{s}}{10\,{\mathrm GeV}}
\right)^{0.4}. 
\]
The singlet-nucleon total cross section $\sigma_{(c\bar{c})_1N}$ is proportional
to the square of the radius $r_{c\bar{c}}$ and has a same energy dependence.
We assume that the transverse distance between the quarks increases linearly
with time until the respective hadronic radius $r_i$ is reached. In the
$c\bar{c}$ rest frame, we write 
\[\label{radius}
r_{c\bar{c}}(\tau) = \left\{ \begin{array}{rl} r_{0}+v_{c\bar{c}}\; \tau &
\qquad {\mathrm if}\ r_{c\bar{c}}(\tau) \leq r_{i},\\ r_{i} \qquad &
\qquad {\mathrm otherwise,}\\ 
\end{array} 
\right. 
\] 
where $i$ stands for $J/\psi$, $\psi'$ and $\chi$. Therefore, the time dependent
cross sections for singlet states are given by 
\[ 
\sigma_{(c\bar{c})_1N}(\tau) =\sigma_{1}\cdot
\left(\frac{\sqrt{s}}{10\,{\mathrm GeV}}\right)^{0.4}
\left(\frac{r_{c\bar{c}}(\tau)}{r_{\psi}}\right)^{2}. 
\label{cross1} 
\]

Using one common set of parameters fitted to E866/NuSea results\footnote{$r_0 = 0.15$~fm, $v_{c\bar{c}}$= 1.85, $\sigma_{1}=2.1$~mb, $\tau_{8\to 1}$= 0.02~fm, $\sigma_{8}=22.3$~mb.}, almost all
other available data on charmonium production in p-A collisions can be
reproduced within this approach~\cite{arl99}.

As this inelastic cross section is intimately related to the time evolution of
the $c\bar{c}$ system that travels through the nucleus, charmonium absorption
within this model is driven by the Lorentz boost $\gamma$ of the $c\bar{c}$ pair
in the nucleus rest frame: 
\[ 
\gamma=\frac{m_{c\bar{c}}}{2 x_2 m_p}, 
\] 
where $x_2$ is the momentum fraction carried by the target parton and
$m_{c\bar{c}}\approx m_{J/\psi}$. It therefore leads to $x_2$ scaling for the nuclear charmonium absorption. 

Let us now turn to the predictions of the model. Using the same set of
parameters, we have calculated charmonium suppression in p(100~GeV)+A(100~AGeV)
collisions over a wide range of rapidities ($-3 \le y_{\mathrm{cm}} \le 3$). The
$J/\psi$ ($dashed$) and $\psi'$ ($dotted$) production ratios $R$(Au/p) are
depicted in Figure~\ref{nuau}. As expected, both ratios dramatically drop around
mid-rapidity. In our model, this fall is a consequence of the increase of
color-octet nucleon inelastic interactions (as compared to the color-singlet nucleon
interactions) with $\gamma$, {\it i.e.} with $x_{F}$. Nevertheless, we would
like to remind the reader that such a qualitative feature can also be seen as a
direct consequence of the $x_2$ scaling assumption, together with the particular
$x_2$ dependence of the E866/NuSea data. Actual {\it predictions} of the model
are in fact the behavior for the rapidity ranges $y_{\mathrm{cm}}<-1$ and
$y_{\mathrm{cm}}> 1$. 

At large $y_{\mathrm{cm}}$ (i.e., large $\gamma$ factor), the partonic
$c\bar{c}$ state travels through the entire nucleus in a color octet state and
hence leads to a constant $R$(Au/p)\footnote{Actually a slight decrease is
expected due to the weak energy dependence of the octet-nucleon cross section
(see section~II.A)} as low as $R \approx 0.2$.

The smaller the  values of $y_{\mathrm{cm}}$ the more likely the color
neutralization to happen inside the nucleus. For negative $y_{\mathrm{cm}}$ the
charmonia have time to  expand to their asymptotic size inside the nucleus.
Then, $J/\psi$ and $\psi'$s have different cross sections with the surrounding
matter and consequently the respective production ratios differ
(Fig.~\ref{nuau}). Finally, let us also mention that these general trends are
found to be similar in lighter systems.

\begin{figure}[htbp] 
\centerline{\psfig{figure=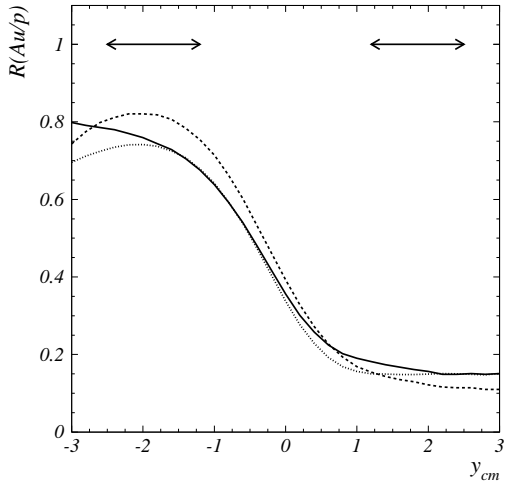,width=7.cm}}
\caption{Ratio R(Au/p) versus $y_{\rm cm}$ of produced $J/\psi$'s in
p(100~GeV)+ Au(100~AGeV) collisions predicted by the model assuming nuclear absorption only ({\it solid}), nuclear absorption with EKS98 shadowing ({\it dashed}), nuclear absorption with BDMPS energy loss ({\it dotted}).}
\label{nu_wl} 
\end{figure}

\section{Shadowing and energy loss} 

So far, charmonium suppression has been attributed to the inelastic interactions of the $\ccb$ pair in nuclear matter. Let us investigate in this section the effects of both shadowing and parton energy loss in the nuclear dependence of charmonium suppression at RHIC energies.

\emph{Shadowing.} It is known from the first EMC data that structure functions $F_2^A(x)$ in a heavy nucleus strongly differ from that measured in light targets. At small $x_2 \ll 1$ (i.e., large $y_{\rm cm}^{\jps}$), the depletion of parton distributions ({\it shadowing}) lowers the charmonium nuclear production, hence leading to a production ratio smaller than one. On the contrary, possible gluon anti-shadowing between $0.1 < x_2 < 0.3$ enhances charmonium production in a nucleus in the range $-3 < y_{\rm cm}^{\jps} < -1.8$ at $\sqrt{s} = 200$~GeV.

Shadowing effects on $\jps$ suppression in proton-nucleus collisions have been investigated by several authors~\cite{shadowing}. They all report that it leads to a $\jps$ suppression at large $x_F$ at Fermilab energy in agreement with the general trend reported by E772~\cite{e772} and E866/NuSea~\cite{e866}.  

We compute in Figure~\ref{nu_wl} ({\it dashed}) the $\jps$ production ratio in $p$(100 GeV)-Au(100 AGeV) assuming both nuclear absorption and shadowing effects. The production cross sections are given by the color evaporation model with MRST parton distributions~\cite{mrst}. Nuclear shadowing is evaluated using the parameterization of deep inelastic scattering and Drell-Yan data by Eskola {\it et al.} (EKS98)~\cite{eks98}. A glimpse at Figure~\ref{nu_wl} indicates that the general trend predicted for the rapidity dependence of $\jps$ suppression still holds when shadowing of target partons is taken into account. We also note that measurements in the negative rapidity range ($-2.2 < y_{\rm cm}^{\jps} < -1.1$) may signal gluon antishadowing effects. Furthermore, charmonium production at large rapidities ($y_{\rm cm}^{\jps} \le 2$) will reveal a possible saturation of shadowing at small $x_2 \sim 10^{-3}$.

While the EKS98 shadowing based on the factorization of the nuclear structure functions leads only to a {\it small} suppression as compared to nuclear absorption, it is worth pointing out that a rapidity dependence for $\jps $ suppression at RHIC similar to the one in Figure~\ref{nu_wl} has also been predicted as coming from coherence effects {\it only}~\cite{Kopeliovich01}.

\emph{Energy loss.} The incoming parton from the projectile may suffer multiple scatterings while traveling through the nucleus $A$, hence leading to a shift of its momentum fraction from $x_1 + \Delta x_1$ to $x_1$ at the time of the $c\bar{c}$ production. In addition to final state interactions, $\psi$ production in a nucleus will thus be suppressed roughly by a factor
\begin{equation}\label{eq:xsprod}
\sigma(p\, p\, \to\, \psi(x_1 + \Delta x_1)\,X)\,/\,\sigma( p\, p\, \to\, \psi(x_1)\,X)
\end{equation}
Because of the dramatic drop of the parton distributions at large $x$ (especially that of the gluon), one may reasonably expect a significant suppression even for small shifts of the momentum fraction $x_1$.

A lot of theoretical effort has been devoted in the last few years to relate the momentum fraction shift $\Delta x_1$ to the atomic mass number $A$ and the center-of-mass energy $\sqrt{s}$ of the collision~\cite{GM,BH,BDMPS}. For the numerical computations to follow, we shall adopt the model by Baier {\it et al.} (BDMPS)~\cite{BDMPS} which assumes the radiative energy loss per unit length $\Delta E/ \Delta z$ to depend linearly on the length $z \propto A^{1/3}$ covered by the parton in the nucleus, i.e.,
\[
\frac{\Delta E}{\Delta z} \propto A^{1/3}
\]
which leads to a momentum fraction shift
\begin{equation}\label{eq:dx1}
\Delta x_1 = \frac{\kappa}{s} A^{2/3}
\end{equation}
where $\kappa$ is a free parameter\footnote{We assume that the perturbative analysis of BDMPS is justified at the energies of RHIC. This may no longer be true at SPS energy where the incoming parton energy is small.}. While the momentum fraction shift (\ref{eq:dx1}) vanishes at large energy $\sqrt{s}$, energy loss effects on charmonium production are expected to be sizeable at low incident energies when $\Delta x_1$ becomes large. We already emphasized that such a mechanism may thus be responsible for the lack of $x_2$ scaling seen from NA3 to Fermilab data in $J/\psi$ production~\cite{moriond00,arleo_these}.

Let us now estimate more quantitatively the effects of energy loss in $J/\psi$ production at RHIC energy. We apply the BDMPS model~\cite{BDMPS} with a quark energy loss $\Delta E / \Delta z =$~0.3 GeV/fm ($\kappa =$~0.086 GeV$^2$) in a gold nucleus\footnote{The corresponding gluon energy loss is expected to be $d{\mathrm E}/dz|_g = 9/4\, d{\mathrm E}/dz|_q \approx$ 0.7 GeV/fm in a gold nucleus.}. Energy loss of this magnitude together with the absorption model appears to be consistent with both NA3 and E866/NuSea $J/\psi$ data~\cite{arleo_these}. Again, the production cross sections in~(\ref{eq:xsprod}) are computed within the color evaporation model and using MRST parton distributions~\cite{mrst}.

The $J/\psi$ production ratio $R$(Au/p) in p(100 GeV)+A(100 AGeV) collisions
including energy loss is displayed in Figure~\ref{nu_wl} ({\it dotted}). Energy loss proves to
be relevant in the nucleus fragmentation region and to be negligible otherwise.
Therefore, the large suppression predicted at RHIC by the $x_2$ scaling assumption
is not altered by such a mechanism. We further note that possible energy loss effects may be seen and better understood~\cite{arleo_these} at HERA-B where charmonium and Drell-Yan production will be measured in
p(920~GeV)-nucleus collisions for $-0.5 \le x_F \le 0.3$~\cite{hera-b}.

Whereas we focus here on hadron-nucleus reactions, we would like to emphasize that parton energy loss in hot QCD matter (or ``jet quenching'',~\cite{wlossQGP}) is expected to be even much larger than that in nuclei of about one order of magnitude~\cite{BDMPS,wlossQGP}. We note that the PHENIX collaboration at RHIC recently reported on the suppression of the yield of high $p_T$ hadrons that may be attributed to finite parton energy loss~\cite{phenixquenching}. In view of that, determination of finite energy loss in proton-nucleus collisions would certainly help to constrain estimates obtained in nucleus-nucleus collisions. Finally, let us also mention that elastic scatterings of the produced $\ccb$ pair with the nucleus ({\it final state} energy loss) may also lead to large effects in the nuclear dependence of $J/\psi$ production~\cite{SatzWL}, although the radiative energy loss of heavy quarks may be much smaller than the one of light quarks~\cite{dokwl}.

Furthermore, we have checked that this particular rapidity dependence is
expected to persist when nuclear shadowing of target partons, predicted by the
EKS98 parameterization~\cite{eks98}, is taken into account.

\section{Experimental considerations} 

The results presented so far show that
charmonium production in p-A collisions at RHIC energies will allow to {\it
discard} experimentally one of the two presently advanced scaling assumptions in
charmonium production. For ruling out one of the assumptions it is indeed
sufficient to measure charmonium production in the rapidity interval
$1<y_{\mathrm{cm}}<2$. Such a measurement can be carried out by the present
setup of the PHENIX experiment which is sensitive to charmonium production in
the dimuon decay channel for the rapidity range $1.2 \le |y_{\rm lab}| \le 2.4$,
represented by black arrows in Figure~\ref{nuau}.

However, if one would like to {\it confirm} scaling, either in $x_F$ or $x_2$, a
precise measurement in the region where R drops sharply is necessary, i.e.
either in the rapidity range $2.6 \le y^{J/\psi}_{\mathrm{cm}}\le 4.1$ or in
$-0.6 \le y^{J/\psi}_{\mathrm{cm}} \le 0.8$ respectively (see
Tables~\ref{table_xf_scaling} and~\ref{table_x2_scaling}, {\it right}). These
domains are most unfortunately not covered by the dimuon spectrometer
acceptance, although the midrapidity region could be explored  via the $e^+e^-$
decay channel of $J/\psi$. 

As for the backward dimuon spectrometer, it will permit to investigate formation
time effects in charmonium production, as demonstrated in Figure~\ref{nuau}. In
particular, the $\psi'$ over $J/\psi$ ratio in various nuclei in this rapidity
range may allow to constrain formation times estimates as well as the time
evolution of the partonic system. Both are not precisely known so far.

\begin{figure}[htbp] 
\centerline{\psfig{figure=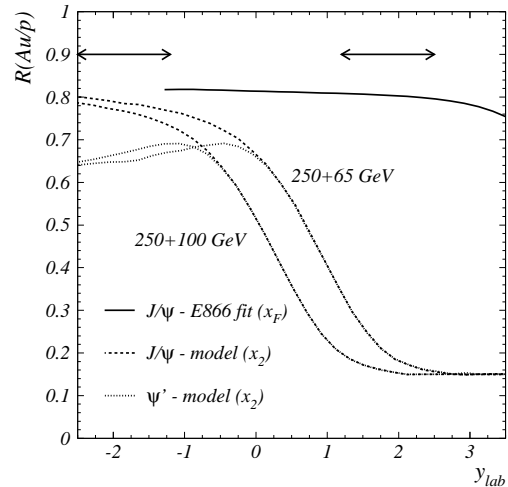,width=7.cm}} 
\caption{Idem as Figure 1 in asymmetric p(250 GeV)+ Au(100 AGeV) and p(250 GeV)+
Au(65 AGeV) collisions.}
\label{asym} 
\end{figure}

One may wonder whether energy asymmetric collisions like p(250~GeV)+A(100~AGeV)
~\cite{twnmtg} allow to investigate scaling with the present detector
acceptance. As shown in Figure~\ref{asym} this is not very promising. In order to observe the steep drop of $R$ in the acceptance region, one has in fact to lower the gold beam energy even further, e.g., p(250 GeV)+Au(65 AGeV).

\section{Baseline to nucleus-nucleus data} 

Heavy quarkonium suppression is
expected to signal a quark-gluon plasma (QGP) formation in nucleus-nucleus
collisions~\cite{mat86}.  On the contrary, no deconfinement is expected in
proton-nucleus collisions. Therefore, p-A measurements provide a  natural
benchmark to which a possible additional suppression due to QGP formation has to
be compared with.

However, we have seen that the expected $J/\psi$ rate in proton-nucleus
collisions is strongly dependent on the scaling law of the underlying mechanism
responsible for charmonium suppression. To illustrate this, let us give an
estimate for $J/\psi$ production ratio in inclusive Au-Au collisions at ${\sqrt
s} = 200$~AGeV, assuming no abnormal suppression (i.e. no QGP formation). In
this case, the production ratio at a given $y_{\rm{cm}}$ is expected to be roughly
\[
R({\mathrm Au}-{\mathrm Au}, y_{{\rm cm}}) 
\approx R({\mathrm Au}-{\mathrm p},y_{{\rm cm}})\,R({\mathrm Au}-{\mathrm p},
-y_{\rm cm})
\]

Assuming $x_{F}$ scaling, we expect a constant production ratio R(Au/p)~$\approx
0.64$ in $y_{\rm cm}\in[-2,2]$. Provided $x_2$ scaling prevails in charmonium
nuclear production, this ratio is also constant and as low as R(Au/p)~$\approx
0.16$.
 
Thus, the extrapolation from presently available p-A data to the hadronic
suppression expected at RHIC energies yields an uncertainty of a factor of four.
Therefore, an unambiguous judgement whether $J/\psi$ suppression seen in A-A
collisions at RHIC is due to QGP formation has to await for p-A measurements at
the same energy over a wide range of rapidities.

\end{document}